\def\year{2019}\relax
\documentclass[letterpaper]{article} %DO NOT CHANGE THIS
\usepackage{aaai19}  %Required
\usepackage{times}  %Required
\usepackage{helvet}  %Required
\usepackage{courier}  %Required
\usepackage{url}  %Required
\usepackage{graphicx}  %Required
\usepackage{bm}
\usepackage{amsmath}
\usepackage{amssymb}
\usepackage{algorithm}
\usepackage{algorithmic}
\usepackage{multirow}
\usepackage{booktabs}
\usepackage{tabularx}
\begin{document}

\newcolumntype{C}[1]{>{\centering\arraybackslash}p{#1}}

% The file aaai.sty is the style file for AAAI Press 
% proceedings, working notes, and technical reports.
%
\title{A Generative Model for Dynamic Networks with Applications}
\author{Shubham Gupta {\normalfont and} Gaurav Sharma {\normalfont and} Ambedkar Dukkipati\\
Department of Computer Science and Automation\\
Indian Institute of Science \\
Bangalore - 560012, India \\
email: \{shubhamg, ambedkar\}@iisc.ac.in
}

\maketitle
\begin{abstract}
Networks observed in real world like social networks, collaboration networks etc., exhibit temporal dynamics, i.e. nodes and edges appear and/or disappear over time. In this paper, we propose a generative, latent space based, statistical model for such networks (called dynamic networks). We consider the case where the number of nodes is fixed, but the presence of edges can vary over time. Our model allows the number of communities in the network to be different at different time steps. We use a neural network based methodology to perform approximate inference in the proposed model and its simplified version. Experiments done on synthetic and real world networks for the task of community detection and link prediction demonstrate the utility and effectiveness of our model as compared to other similar existing approaches.
\end{abstract}

%========================================================================================================================%

\section{Introduction}
\label{section:introduction}

Many networks encountered in real world evolve with time, i.e. new nodes and edges appear while some existing ones disappear. For example, consider the network of people connected via Facebook, where the users and their interactions change over time. It is interesting to study the underlying dynamics of such networks to understand what drives these changes, how communities are formed, how will a network behave in future and what has lead to the current state of the network etc.

In static network setting, one of the problems that has been extensively studied is community detection \cite{Fortunato:2010:CommunityDetectionInGraphs}. In a dynamic setting, this problem becomes more challenging since communities themselves take birth, meet death, grow, shrink, split and merge etc. \cite{RossettiEtAl:2017:CommunityDiscoverInDynamicNetworksASurvey}. In this paper,  we address the problem of modeling dynamic networks that exhibit a community structure where the number of nodes is invariant over time but the presence of edges is time dependent.

A wide range of dynamic networks with different characteristics can be found in real world. In order to model these networks in a meaningful way, it is essential to focus on certain specific types of networks. In this paper, we consider undirected (edges are not directional), assortative (similar types of nodes have a high probability of connecting), positive influence based (adjacent nodes positively influence a given node) and gradually evolving networks. Many real world networks such as friendship networks, communication networks etc., fall in this category.

We propose a generative model, which we call Evolving Latent Space Model (ELSM), for dynamic networks of the type mentioned above. The main advantage of our model over existing models \cite{XingEtAl:2010:AStateSpaceMixedMembershipBlockmodelForDynamicNetworkTomography,FouldsEtAl:2011:ADynamicRelationalInfiniteFeatureModelForLongitudinalSocialNetworks,HeaukulaniEtAl:2013:DynamicProbabilisticModelsForLatentFeaturePropagationInSocialNetworks,KimEtAl:2013:NonparametricMultiGroupMembershipModelForDynamicNetworks,XuHero:2014:DynamicStochasticBlockmodelsForTimeEvolvingSocialNetworks}, is that our model uses a more flexible, continuous latent space and we allow the number of communities to vary over time (also supported by \cite{KimEtAl:2013:NonparametricMultiGroupMembershipModelForDynamicNetworks}). It also allows one to generate a dynamic graph with a temporally \textit{stable} community structure. The purpose of proposing this generative model is three-folds: (i) it will help in understanding how networks evolve over time, (ii) it will provide synthetic data for other dynamic community detection and link prediction algorithms to operate on and (iii) it will allow model based inference for problems like community detection and link prediction.

Though ELSM can be used both as a generative model and an inference model, for the inference problem, we propose a simplified version of our model that is computationally less expensive. This model can be used exclusively for inference and we refer to it as iELSM. Inference in ELSM is given in supplementary material \cite{GuptaEtAl:2018:AGenerativeModelForDynamicNetworksWithApplicationsSupplementaryMaterial}. Exact inference in ELSM and iELSM is intractable, thus we resort to variational techniques for performing approximate inference. We use a recurrent neural network (RNN)  to model various variational parameters. We also present an extension of our model that deals with weighted graphs in the experiments section (\S \ref{section:experiments}).

Our main contributions are:
\textbf{(i)} We propose a generative model (ELSM) for dynamic networks with fixed number of nodes that allows the number of communities to vary over time.
\textbf{(ii)} We derive a variational approximation to exact inference in ELSM and iELSM and use a neural network architecture to learn variational parameters.
\textbf{(iii)} We outline a general inference methodology that can be customized in various ways by using problem specific information.
\textbf{(iv)} We demonstrate the applicability of proposed model and inference methodology by performing experiments on real and synthetic networks for community detection and link prediction where we obtain state of the art results.

%%%%%%%%%%%%%%%%%%%%%%%%%%%%%%%%%%%%%%%%%%%%%%%%%%%%%%%%%%%%%%%%%%%%%%%%%%%%%%%%%%%%%%%%%%%%%%%%%%%%%%%%%%%%%%%%%%%%%%%%%%%%%%

\section{Evolving Latent Space Model}
\label{section:ELSM}
\textbf{Notation: }We use capital bold letters ($\mathbf{A}$, $\bm{\Sigma}$ etc.) for matrices and set of vectors, small bold letters ($\mathbf{m}$, $\bm{\alpha}$ etc.) for vectors and small letters ($a$, $b$ etc.) for scalars. If a quantity is associated with time step $t$, we add $(t)$ to the superscript as in $\mathbf{A}^{(t)}$. We use the same letter for denoting set of vectors and an element in this set. For example, $\bm{\Sigma}^{(t)}$ is a set of vectors associated with time step $t$ and $\bm{\sigma}_i^{(t)}$ is the $i^{th}$ vector in this set. The $j^{th}$ element of a vector $\mathbf{c}$ is denoted by $c_j$ (note that this is not bold).

Our model draws motivation from two phenomena that are frequently observed in real world networks: (i) \textit{similar} nodes are more likely to connect to each other and (ii) over a time, nodes tend to become similar to their neighbors. We aim to model a latent space, in which the euclidean distance between embeddings of any two nodes is inversely proportional to their similarity. In this space, operations mimicking the two phenomena listed above can be modeled naturally as we explain later in this section.

To make the underlying motivations concrete, our running example in this section would pertain to the political inclinations of $n$ hypothetical individuals. In this context, the latent space can be thought of as an ideological space where the axes correspond to different political ideologies. Each individual can be embedded in this space based on their inclination towards different ideologies. Distance between two latent embeddings will then correspond to difference in political standpoints of the corresponding individuals. The observations will be symmetric, binary adjacency matrices $\mathbf{\{A}^{(t)}\}_{t=1}^T$ where $a_{ij}^{(t)} = 1$ if individuals $i$ and $j$ interacted with each other in time window $t$. One can see that the two phenomena listed in the previous paragraph naturally occur in this setting, i.e., individuals tend to interact with people who share their ideology and over time the peer group of an individual positively influences their ideology.

Our proposed generative model can be summarized as follows: first, a set of initial latent embeddings, $\mathbf{Z}^{(1)} =\{\mathbf{z}_1^{(1)}, \mathbf{z}_2^{(1)}, ..., \mathbf{z}_n^{(1)}\}$, is created. Based on the latent embeddings $\mathbf{Z}^{(1)}$ the observed adjacency matrix $\mathbf{A}^{(1)}$ is sampled. For $t$ = 2, 3 ... $T$, latent embeddings at time $t$, $\mathbf{Z}^{(t)}$, are updated based on $\mathbf{Z}^{(t-1)}$, $\mathbf{A}^{(t-1)}$ and some extra information about possible new communities at time-step $t$ and $\mathbf{A}^{(t)}$ is sampled based on $\mathbf{Z}^{(t)}$. Next, we describe each component of this process in detail.

%============================================================================================================================
\begin{figure}
\begin{center}
\centering
\includegraphics[width=\columnwidth]{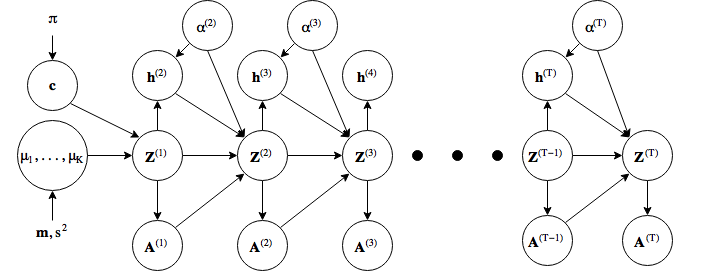}
\caption{Graphical Model for ELSM. $h^{(t)}_i = 1$ for nodes that join the newly created community with center $\bm{\alpha}^{(t)}$ and $0$ otherwise. This \textit{splitting} mechanism models the birth of new communities.}
\label{fig:graphicalmodel}
\end{center}
\end{figure}

\subsection{Getting the Initial Set of Embeddings}
\label{section:gettinginitialsetofembeddings}
Let $K$ denote the maximum number of clusters at time $t$ = 0. We take $K$ and a multinomial distribution over $K$ elements, $\mathbf{\pi}$, as input. A vector $\mathbf{c} \in \mathbb{R}^n$ is created such that it assigns an integer between $1$ and $K$ sampled from $\pi$ to each node . In the general case, we sample the initial community centers $\bm{\mu}_1, \bm{\mu}_2, ..., \bm{\mu}_K \sim \mathcal{N}(\mathbf{m},\, s^2\mathbf{I})$, where $\mathbf{m} \in \mathbb{R}^d$ and $s^2 \in \mathbb{R}$ are hyperparameters and $d$ is the dimension of latent space. The embedding for each node in the first snapshot can then be sampled as:
\begin{equation}
    \label{eq:initial-assignment}
    \mathbf{z}^{(1)}_i \sim \mathcal{N}(\bm{\mu}_{c_i},\, s_1^2\mathbf{I}), \,\,\,\, i=1, 2, ..., {n}
\end{equation}
The parameter $s_1$ dictates the \textit{spread} within communities. In the context of our example, this step corresponds to the creation of an initial set of political ideologies and individuals subscribing to these ideologies. Since $\bm{\mu}_1, \bm{\mu}_2, ..., \bm{\mu}_K$ are sampled independently, there will be at-most $K$ distinct communities in latent space for the first time step. To encode prior information about communities in the model, rather than sampling $\bm{\mu}_1, \bm{\mu}_2, ..., \bm{\mu}_K$ independently from a normal distribution, one can specify these vectors explicitly.

%============================================================================================================================

\subsection{Generating  Observed $\mathbf{A}^{(t)}$ from Latent $\mathbf{Z}^{(t)}$}
\label{section:generatingobservedatfromlatentzt}
In ELSM, the probability of an edge between two nodes is inversely proportional to the distance between their embeddings in the latent space. By this, one hopes to get embeddings that are easier to deal with for a clustering algorithm like \textit{k-means}. We model this as follows:
\begin{equation}
    \label{eq:observedgivenlatent}
    a_{ij}^{(t)} = a_{ji}^{(t)} \sim \mathrm{Bernoulli}\Big(f(\mathbf{z}_i^{(t)} - \mathbf{z}_j^{(t)})\Big), \,\,\,\, i > j,
\end{equation}
where $f(.)$ is a nonlinear function. The observed matrices are symmetric and we do not allow self loops. In our experiments on generation of synthetic networks, we used $f(x) = 1 - \mathrm{tanh}(||x||_2^2/{s}_2^2)$, where ${s}_2$ is a parameter that controls the radius around a node's embedding in the latent space within which it is more likely to connect to other nodes.

We will demonstrate via experiments in \S \ref{section:experiments}, that one can have other choices for the function $f(.)$ and the distribution parameterized by $f(.)$ depending on problem context. For example, in the case of weighted graph with positive integer weights, Poisson distribution can be used. 

%============================================================================================================================

\subsection{Emergence of New Communities}
\label{section:emergenceofnewcommunities}
If the nodes are only allowed to become similar to their neighbors, then over time the latent embeddings will collapse onto a single point and all the community structure in the network will be lost. But such a phenomenon is not observed in real world networks. In the absence of any other guiding force, to avoid the collapse of latent space and to model the emergence of new communities ELSM randomly generates a new community center at each time step.

At each time step $t$, a new community center $\bm{\alpha}^{(t)}$ is sampled from the same prior that was used in the first layer to generate the original community centers $\bm{\mu}_1, \bm{\mu}_2, ..., \bm{\mu}_K$, i.e. $\bm{\alpha}^{(t)} \sim \mathcal{N}(\mathbf{m},\, s^2\mathbf{I})$. We define a Bernoulli random variable ${h_i^{(t)}}$ for each node ${i}$ at time ${t}$. If ${h_i^{(t)}} = 1$, then node ${i}$'s updated latent embedding is sampled as:
\begin{equation}
    \label{eq:splitting}
    \mathbf{z}_i^{(t)} \sim \mathcal{N}(\bm{\alpha}^{(t)}, s_1^2\mathbf{I})
\end{equation}
The parameter ${s_1}$ has the same meaning as in (\ref{eq:initial-assignment}). If ${h_i^{(t)}} = 0$, then node ${i}$'s latent embedding will be updated based on the embeddings of its neighbors at time step ${t-1}$ by the process described in the next subsection. We model the probability $P({h_i^{(t)}} = 1)$ as a function of the distance between $\bm{\alpha}^{(t)}$ and $\mathbf{z}_i^{(t-1)}$. If the latent embedding of node ${i}$ at time step ${t-1}$ is close to $\bm{\alpha}^{(t)}$, then $P({h_i^{(t)}} = 1)$ will be high and vice-versa. We model this as follows:
\begin{equation}
    \label{eq:changeprobability}
    {h_i^{(t)}} \sim \mathrm{Bernoulli}\Big(g(\mathbf{z}_i^{(t-1)} - \bm{\alpha}^{(t)})\Big), 
\end{equation}
where $g(.)$ is a nonlinear function. In our experiments for generating synthetic data, we use $g(x) = 1 - \mathrm{tanh}(||x||_2^2/{s_3^2})$. The parameter ${s_3}$ controls the influence of the newly created community on other nodes. Note that this step does not necessarily involve the creation of a new community as the sampled $\bm{\alpha}^{(t)}$ may lie within an existing community in the latent space.

In the context of our example, this step will correspond to new political ideologies appearing in the latent space. Individuals that have a similar political ideology are more like to embrace this new ideology. Individuals from different communities may come together and form a new political ideology of their own.

%============================================================================================================================

\subsection{Evolving the Latent Space}
\label{section:evolvingthelatentspace}
ELSM tries to make nodes similar to their neighbors over time. To model this, for time step ${t} = 2, 3, ..., {T}$, a mean vector $\bm{\mu}_i^{(t)}$ is obtained for each node ${i}$ as follows:
\begin{equation}
\label{eq:latentupdatemean}
\begin{split}
    \bm{\mu}_i^{(t)} =& \frac{1}{1 + \sum_{j \neq i} {a_{ij}^{(t-1)}} l(\mathbf{z}_i^{(t-1)} - \mathbf{z}_j^{(t-1)})} \times    \Big(\mathbf{z}_i^{(t-1)} \\
    &+ \sum_{j \neq i} {a_{ij}^{(t-1)}} l(\mathbf{z}_i^{(t-1)} - \mathbf{z}_j^{(t-1)}) \mathbf{z}_j^{(t-1)}\Big)
\end{split}
\end{equation}
Here, we use $l(x) = e^{-||x||^2/{s_4^2}}$ and ${s_4}$ is a parameter that controls the influence of neighbors on a node. Note that only the immediate neighbors at time ${t-1}$ can affect the future embedding of a node at time ${t}$. Also, note that the value of $\bm{\mu}_i^{(t)}$ lies in the convex hull formed by the embeddings of the node and its neighbors in the latent space a time step $t-1$. If ${h_i^{(t)}} = 0$ for a given node, then the updated embedding of the node is given by:
\begin{equation}
    \label{eq:updatedembedding}
    \mathbf{z}_i^{(t)} \sim \mathcal{N}(\bm{\mu}_i^{(t)}, {s_1^2}\mathbf{I})
\end{equation}
Then, (\ref{eq:splitting}) and (\ref{eq:updatedembedding}) can be combined to get the general update equation for the evolution of latent embeddings of the nodes as:
\begin{equation}
    \label{eq:generalizedupdatedembedding}
    \mathbf{z}_i^{(t)} \sim \mathcal{N}({h_i^{(t)}}\bm{\alpha}^{(t)} + (1-{h_i^{(t)}})\bm{\mu}_i^{(t)}, {s_1^2}\mathbf{I}), \,\,\,\, {i} = 1, 2, ..., {n}
\end{equation}
This not only allows the nodes to become similar over time, it also allows them to move apart to form new communities as it is observed in real world networks. In the context of our example, (\ref{eq:latentupdatemean}) says that an individual's political ideology will be influenced by the political ideology of their peers.

It is easy to see that ELSM supports operations like birth, death, growth, shrinkage, merge and split on the communities. The latent space offers a nice community structure (as we will show via our experiments, visualization of latent space is in supplementary material) because the operations presented in (\ref{eq:observedgivenlatent}), (\ref{eq:changeprobability}) and (\ref{eq:latentupdatemean}) depend on distances between embeddings in the latent space. Figure \ref{fig:graphicalmodel} presents the underlying graphical model for ELSM. The procedure for generating synthetic networks is listed in Algorithm \ref{alg:generatingsyntheticdata}.

\begin{algorithm}[tb]
   \caption{Generating Synthetic Networks}
   \label{alg:generatingsyntheticdata}
\begin{algorithmic}
   \STATE {\bfseries Input:} ${n}$, ${T}$, $\mathbf{\pi}$, ${K}$, $\mathbf{m}$, ${s}$, ${s_1}$, ${s_2}$, ${s_3}$, ${s_4}$
   \STATE Sample $\bm{\mu}_1$, $\bm{\mu}_2$, ..., $\bm{\mu}_K \sim \mathcal{N}(\mathbf{m}, {s^2}\mathbf{I})$
   \STATE Sample ${c_1}$, ${c_2}$, ..., ${c_n} \sim \mathbf{\pi}$
   \STATE Sample $\mathbf{Z}^{(1)}$ using (\ref{eq:initial-assignment})
   \STATE Sample $\mathbf{A}^{(1)}$ using (\ref{eq:observedgivenlatent})
   \FOR{$t=2$ {\bfseries to} ${T}$}
        \STATE Sample $\bm{\alpha}^{(t)} \sim \mathcal{N}(\mathbf{m}, {s^2}\mathbf{I})$
        \STATE Sample ${h^{(t)}}$ using (\ref{eq:changeprobability})
        \STATE Sample $\mathbf{Z}^{(t)}$ using (\ref{eq:generalizedupdatedembedding})
   \STATE Sample $\mathbf{A}^{(t)}$ using (\ref{eq:observedgivenlatent})
   \ENDFOR
   \STATE {\bfseries Return:} $\{\mathbf{A}^{(1)}, \mathbf{A}^{(2)}, ..., \mathbf{A}^{(T)}\}$
\end{algorithmic}
\end{algorithm}

%%%%%%%%%%%%%%%%%%%%%%%%%%%%%%%%%%%%%%%%%%%%%%%%%%%%%%%%%%%%%%%%%%%%%%%%%%%%%%%%%%%%%%%%%%%%%%%%%%%%%%%%%%%%%%%%%%%%%%%%%%%%%%

\section{Inference Network}
\label{section:inferenceinlatentevolutionmodel}

\begin{figure}
\begin{center}
\centering
\includegraphics[scale=0.35]{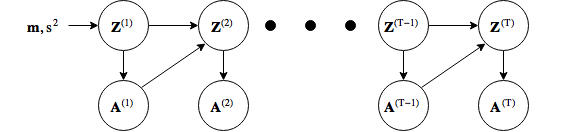}
\caption{Graphical Model for iELSM}
\label{fig:graphicalmodelsELSM}
\end{center}
\end{figure}

Inference is concerned with finding the latent embedding of each node at each time step, $\{\mathbf{Z}^{(1)}, \mathbf{Z}^{(2)}, ..., \mathbf{Z}^{(T)}\}$, that best explains an observed dynamic network specified by the sequence of adjacency matrices $\{\mathbf{A}^{(1)}, \mathbf{A}^{(2)}, ..., \mathbf{A}^{(T)}\}$. In this section, we describe the inference procedure for a simplified version of ELSM. We argue that this simplified version, which we call Evolving Latent Space Model for Inference (iELSM), is as good as the original ELSM for the task of inference. Inference for ELSM follows along the same lines as iELSM and full details have been worked out in supplementary material.

%============================================================================================================================

\subsection{Evolving Latent Space Model for Inference}
\label{section:sELSM}
While generating a synthetic dynamic network using ELSM, a guiding force is needed to: (i) select the initial community membership for each node so that there is a community structure in the first layer and (ii) force nodes to move apart over time and form new communities. In the absence of any other guiding force during the generation process, ELSM achieves this by the use of initial community centers $\bm{\mu}_1, \bm{\mu}_2, ..., \bm{\mu}_K$, the initial membership vector $\mathbf{c}$ and the splitting mechanism as explained in \S \ref{section:emergenceofnewcommunities}. 

However, during inference, given a sequence of  observed adjacency matrices $\{\mathbf{A}^{(1)}, \mathbf{A}^{(2)}, ..., \mathbf{A}^{(T)}\}$ for a dynamic network, the required guiding force is provided by the need to explain the observed data. In this case one need not incorporate any of the elements mentioned above. Thus for the purpose of inference, we eliminate these elements and get a simplified model which we call iELSM. The graphical model for iELSM is shown in Figure \ref{fig:graphicalmodelsELSM}. Note that iELSM also captures both the motivating phenomena (similar nodes connect more often and nodes become similar to their neighbors over time) mentioned in the beginning of \S \ref{section:ELSM}.

The structures of probability distributions that govern iELSM are similar to the structures of corresponding distributions in ELSM. The latent embeddings $\mathbf{z}_i^{(t)}$ follow (\ref{eq:latentupdatemean}) and (\ref{eq:updatedembedding}) for ${i} = 1, 2, ..., {n}$ and ${t} = 2, 3, ..., {T}$. The observed entries ${a_{ij}^{(t)}}$ follow (\ref{eq:observedgivenlatent}). In addition we impose a prior on the first layer of latent embeddings $\mathbf{z}_i^{(1)}$ for ${i} = 1, 2, ..., {n}$ as:
\begin{equation}
    \label{eq:firstlayerprior}
    \mathbf{z}_i^{(1)} \sim \mathcal{N}(\mathbf{m}, {s^2}\mathbf{I}),
\end{equation}
Where $\mathbf{m}$ and ${s}$ are hyperparameters. We make the following independence assumptions: 
\begin{enumerate}
    \item The latent embeddings in the first layer $\mathbf{z}_i^{(1)}$ for ${i} = 1, 2, ..., {n}$ are independent.
    \item ${a_{ij}^{(t)}}$ is independent of everything else given $\mathbf{z}_i^{(t)}$ and $\mathbf{z}_j^{(t)}$ for all ${i}, {j} = 1, 2, ..., {n}, i \neq j$ and ${t} = 1, 2, ..., {T}$.
    \item $\mathbf{z}_{i}^{(t)}$ is independent of everything else given $\mathbf{Z}^{(t-1)}$ and $\mathbf{A}^{(t-1)}$ for ${i} = 1, 2, ..., {n}$ and ${t} = 2, 3, ..., {T}$
\end{enumerate}
Since the observed adjacency matrices are assumed to be symmetric and self-loops are not allowed, we only need to consider the lower triangle (without the leading diagonal) in these matrices. The joint log likelihood for iELSM, which is given below, can then be computed using (\ref{eq:observedgivenlatent}), (\ref{eq:latentupdatemean}), (\ref{eq:updatedembedding}) and (\ref{eq:firstlayerprior}) as:
\begin{equation}
    \label{eq:sELSMjointloglikelihood}
    \begin{split}
    \log P(&\mathbf{A}^{(1)}, \mathbf{Z}^{(1)}, \mathbf{A}^{(2)}, \mathbf{Z}^{(2)}, ..., \mathbf{A}^{(T)}, \mathbf{Z}^{(T)}) = \\
    &\sum_{i=1}^{n}\log P(\mathbf{z}_i^{(1)}) + \sum_{t=1}^{T} \sum_{i > j} \log P({a_{ij}^{(t)}} | \mathbf{z}_{i}^{(t)}, \mathbf{z}_{j}^{(t)}) \\
    & + \sum_{t=2}^{T} \sum_{i=1}^{n} \log P(\mathbf{z}_{i}^{(t)} | \mathbf{Z}^{(t-1)}, \mathbf{A}^{(t-1)})
    \end{split}
\end{equation}

%============================================================================================================================

\subsection{Variational Inference for iELSM}
\label{section:variationalinferenceforsELSM}
Exact inference is intractable in iELSM because the observed entries of $\mathbf{A}^{(t)}$ do not follow a Gaussian distribution. Thus, we resort to variational techniques \cite{BleiEtAl:2017:VariationalInferenceAReviewForStatisticians} to perform approximate inference. We use a LSTM based neural network to model the variational parameters. The neural network is then trained to maximize the $\mathrm{ELBO}$ which we derive next.

We approximate the true posterior distribution $P(\mathbf{Z}^{(1)}, \mathbf{Z}^{(2)}, ..., \mathbf{Z}^{(T)} | \mathbf{A}^{(1)}, \mathbf{A}^{(2)}, ..., \mathbf{A}^{(T)})$ using a distribution $Q(\mathbf{Z}^{(1)}, \mathbf{Z}^{(2)}, ..., \mathbf{Z}^{(T)}; \theta)$, where $\theta$ represents parameters of a neural network. We further assume that $Q(., \theta)$ belongs to the mean-field variational family of distributions, i.e.:
\begin{equation}
    \label{eq:meanfield}
    Q(\mathbf{Z}^{(1)}, \mathbf{Z}^{(2)}, ..., \mathbf{Z}^{(T)}; \theta) = \prod_{t = 1}^{T} \prod_{i=1}^{{n}} q_{ti}(\mathbf{z}_i^{(t)}; \theta)
\end{equation}
We model each factor $q_{ti}(\mathbf{z}_i^{(t)}; \theta)$ as a Gaussian distribution whose mean $\bm{\nu}_i^{(t)} \in \mathbb{R}^d$ and variance ${(\bm{\sigma}_i^{(t)})^2} \in \mathbb{R}^d$ parameters are derived from the neural network. Here, as before, $d$ is the dimension of latent space. Note that we have not explicitly shown the dependence of $\bm{\nu}_i^{(t)}$ and ${(\bm{\sigma}_i^{(t)})^2}$ on $\theta$ to keep the notation clean. Let ${(\bm{\sigma}_i^{(t)})^2} \mathbf{I} = diag((\sigma_i^{(t)})_1^2, (\sigma_i^{(t)})_2^2, ..., (\sigma_i^{(t)})_d^2)$, we can then write:
\begin{equation}
    \label{eq:meanfieldfactordefinition}
    q_{ti}(\mathbf{z}_i^{(t)}; \theta) = \mathcal{N}(\mathbf{z}_i^{(t)} \,\,|\,\, \bm{\nu}_i^{(t)},  {(\bm{\sigma}_i^{(t)})^2} \mathbf{I})
\end{equation}
The objective is to maximize the $\mathrm{ELBO}$ function which is given by $\mathbb{E}_{\mathbf{z} \sim q}[\log p(\mathbf{x}, \mathbf{z}) - \log q(\mathbf{z})]$ \cite{BleiEtAl:2017:VariationalInferenceAReviewForStatisticians}. Here $\mathbf{x}$ denotes the observed variables, $\mathbf{z}$ denotes the latent variables, $p(\mathbf{x}, \mathbf{z})$ is the joint distribution of $\mathbf{x}$ and $\mathbf{z}$ and $q(\mathbf{z})$ is the distribution that approximates the posterior $p(\mathbf{z}|\mathbf{x})$. It can be shown that $\mathrm{ELBO}$ provides a lower bound on the log probability of observed data $p(\mathbf{x})$ \cite{KingmaEtAl:2013:AutoEncodingVariationalBayes}, hence $\mathrm{ELBO}$ can be used as a surrogate function to maximize the probability of observed data. In our context, $\mathrm{ELBO}$ can be written as:
\begin{equation}
    \label{eq:elbo}
    \begin{split}
        \mathrm{ELBO}(\theta) = &\mathbb{E}_{\mathbf{Z}^{(1)}, \mathbf{Z}^{(2)}, ..., \mathbf{Z}^{(T)} \sim Q}[ \\ 
        &\log P(\mathbf{A}^{(1)}, \mathbf{Z}^{(1)}, ..., \mathbf{A}^{(T)}, \mathbf{Z}^{(T)}) \\
        &- \log Q(\mathbf{Z}^{(1)}, \mathbf{Z}^{(2)}, ..., \mathbf{Z}^{(T)}; \theta)]    
    \end{split}
\end{equation}
The expectation of second term in (\ref{eq:elbo}) can be computed in closed form since $Q(., \theta)$ factorizes according to (\ref{eq:meanfield}) and the factors $q_{ti}(.; \theta)$ follow a Gaussian distribution. To compute the expectation of first term, we use the Monte Carlo method to evaluate (\ref{eq:sELSMjointloglikelihood}). In all our experiments, only one sample was used to approximate the expectation of first term.

%============================================================================================================================

\subsection{Network Architecture}
\label{section:networkarchitecture}

\begin{figure}
\begin{center}
\centering
\includegraphics[scale=0.35]{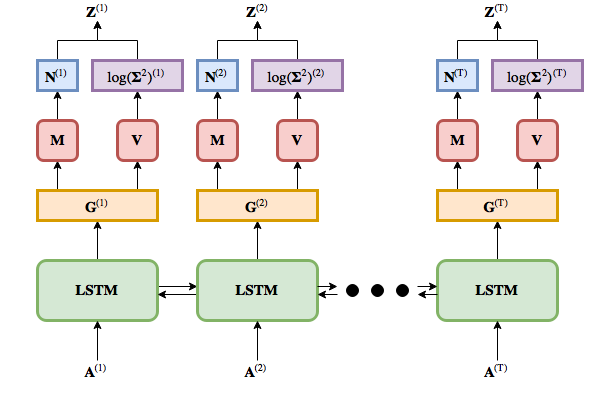}
\caption{Inference network architecture for iELSM. $\mathbf{M}$ and $\mathbf{V}$ are neural networks that take $\mathbf{g}_i^{(t)}$ as input to produce the mean and $\log$-variance vectors respectively. Note that these networks are shared across all timesteps.}
\label{fig:networkmodel}
\end{center}
\end{figure}

Given the variational parameters $\bm{\nu}_i^{(t)}$ and ${(\bm{\sigma}_i^{(t)})^2}$ for all ${i} = 1, 2, ..., {n}$ and ${t} = 1, 2, ..., {T}$, $\mathrm{ELBO}$ can be approximated using (\ref{eq:elbo}) as described above. In this section we describe the neural network architecture that parameterizes these quantities. There are two main advantages of this neural network based approach: (i) neural networks are powerful function approximates and hence complex node to latent embedding mappings can be learned and (ii) standard backpropagation algorithm with the reparameterization trick \cite{KingmaEtAl:2013:AutoEncodingVariationalBayes} can be used that allows efficient implementation.

The recurrent neural network is a natural model for sequential data. We use a bidirectional LSTM in our inference network. At time step ${t}$, our network takes the adjacency matrix $\mathbf{A}^{(t)}$ as input. Each row is treated as an input feature vector. The network updates its hidden state using the usual update rules for LSTM and produces an output $\mathbf{g}_i^{(t)} \in \mathbb{R}^{m}$ for each node. The output $\mathbf{g}_i^{(t)}$ is then used to produce $\bm{\nu}_i^{(t)}$ and $\log {(\bm{\sigma}_i^{(t)})^2}$ via the mean and variance networks (${\mathbf{M}}$ and ${\mathbf{V}}$ respectively in Figure \ref{fig:networkmodel}) that are shared across all time steps. Figure \ref{fig:networkmodel} shows the architecture of our inference network.

We use the reparameterization trick for Gaussian distribution, as it was used in \cite{KingmaEtAl:2013:AutoEncodingVariationalBayes} to make the whole pipeline differentiable. This enables end to end training of the network. The embeddings sampled from this network can be used to compute ({\ref{eq:elbo}}) via Monte Carlo approximation as it was discussed in \S \ref{section:variationalinferenceforsELSM}. The objective is to maximize $\mathrm{ELBO}(\theta)$ with respect to the parameters $\theta$ of the inference network.

%============================================================================================================================

\subsection{Encoder-Decoder view of the Inference Network}
\label{section:encoderdecoderviewofinferencenetwork}
The network given in Figure \ref{fig:networkmodel} can be seen as an encoder. At time step ${t}$, it takes the feature vector for each node (the row corresponding to that node in $\mathbf{A}^{(t)}$) as input and produces a time dependent embedding $\mathbf{z}_i^{(t)}$ for the node as output. Note that if observed node features are available, one can use these features in conjunction with $\mathbf{A}^{(t)}$ as input to the network. A decoder should take these latent embeddings and try to reconstruct $\mathbf{A}^{(t)}$. By using (\ref{eq:sELSMjointloglikelihood}) in (\ref{eq:elbo}), we obtain a term that involves $\log P({a_{ij}^{(t)}} | \mathbf{z}_{i}^{(t)}, \mathbf{z}_{j}^{(t)})$. The estimator of this probability value acts as a simple decoder that tries to predict the probability of an observed edge given the latent embeddings of its two endpoints.

Recall from (\ref{eq:observedgivenlatent}) that: (i) ${a_{ij}^{(t)}}$ has been modeled by a Bernoulli distribution and (ii) The parameter for this Bernoulli distribution is given by $f(\mathbf{z}_i^{(t)} - \mathbf{z}_j^{(t)})$ where for generation we use $f(x) = 1 - \mathrm{tanh}(||x||_2^2/{s_2^2})$. One can have a more complex decoder which utilizes problem specific information. For example, the function $f(x)$ can be modeled using a neural network. If the weights do not follow the Bernoulli distribution (i.e., the edges are weighted) then one can use the problem context to enforce the right distribution on weights and learn the parameters of that distribution accordingly. For example, in our experiments with Enron dataset, the edge weights are non-negative integers, so we model this by using a Poisson distribution and we use a neural network to learn $f(x)$ so that it predicts the mean for the Poisson distribution.

This encoder-decoder view of the inference network makes the inference methodology very flexible. One can incorporate domain knowledge to customize various components of this network so that it better suits the problem at hand.

%%%%%%%%%%%%%%%%%%%%%%%%%%%%%%%%%%%%%%%%%%%%%%%%%%%%%%%%%%%%%%%%%%%%%%%%%%%%%%%%%%%%%%%%%%%%%%%%%%%%%%%%%%%%%%%%%%%%%%%%%%%%%%

\section{Related Work}
\label{section:relatedwork}
In general, dynamic networks, and in particular, link prediction and community detection in dynamic networks, have earned the attention of many researchers \cite{GolderbergEtAl:2010:ASurveyOfStatisticalNetworkModels,KimEtAl:2017:AReviewOfDynamicNetworkModelsWithLatentVariables}. This can be attributed to the emergence of many practical applications (for e.g., social network analysis) where dynamic networks are encountered.

An extension of Mixed Membership Stochastic Block Model \cite{Airoldi:2008:MixedMembershipStochasticBlockmodels} to the dynamic network setting by coupling it with a state-space model to capture temporal dynamics has been proposed in \cite{XingEtAl:2010:AStateSpaceMixedMembershipBlockmodelForDynamicNetworkTomography,HoEtAl:2011:EvolvingClusterMixedMembershipBlockmodelforTimeVaryingNetworks}. Along the same lines, in \cite{YangEtAl:2011:DetectingCommunitiesAndTheirEvolutionsInDynamicSocialNetworksABayesianapproach}, Stochastic Block Model \cite{HollandEtAl:1983:StochasticBlockmodelsFirstSteps} was extended by explicitly modeling transition between communities over time. Some other approaches that extend a static network models are \cite{XuHero:2014:DynamicStochasticBlockmodelsForTimeEvolvingSocialNetworks,Xu:2015:StochasticBlockTransitionModelsForDynamicNetworks,PapadopoulosEtAl:2012:PopularityVsSimilarityInGrowingNetworks}.

There have been many latent space based approaches for modeling dynamic networks. In \cite{SwellEtAl:2015:LatentSpaceModelsForDynamicNetworks,SwellEtAl:2016:LatentSpaceModelsForDynamicNetworksWithWeightedEdges} the transition of nodes' latent embeddings is modeled independently for each node by using a Markov chain. In \cite{HeaukulaniEtAl:2013:DynamicProbabilisticModelsForLatentFeaturePropagationInSocialNetworks}, the authors introduced Latent Feature Propagation (LFP) that uses a discrete latent space and a HMM style latent embedding transition model. In \cite{KimEtAl:2013:NonparametricMultiGroupMembershipModelForDynamicNetworks} non-parametric Dynamic Multigroup Membership Graph model (DMMG) which additionally models the birth and death of communities using Indian Buffet Process \cite{Griffiths:2011:TheIndianBuffetProcessAnIntroductionAndReview} has been introduced. The work of \cite{MillerEtAl:2009:NonparametricLatentFeatureModelsForLinkPrediction,FouldsEtAl:2011:ADynamicRelationalInfiniteFeatureModelForLongitudinalSocialNetworks} are also along the same lines.

Most similar to our work, is the work done in \cite{HeaukulaniEtAl:2013:DynamicProbabilisticModelsForLatentFeaturePropagationInSocialNetworks,KimEtAl:2013:NonparametricMultiGroupMembershipModelForDynamicNetworks}, however there are significant differences as well. (i) While these approaches rely on MCMC, we use variational inference because it offers many advantages as mentioned in \cite{BleiEtAl:2017:VariationalInferenceAReviewForStatisticians}, (ii) We use a neural network to perform inference which allows an efficient implementation and (iii) We have a more flexible, continuous latent space whereas these approaches use a discrete latent space.

%%%%%%%%%%%%%%%%%%%%%%%%%%%%%%%%%%%%%%%%%%%%%%%%%%%%%%%%%%%%%%%%%%%%%%%%%%%%%%%%%%%%%%%%%%%%%%%%%%%%%%%%%%%%%%%%%%%%%%%%%%%%%%

\section{Experiments}
\label{section:experiments}
In this section we demonstrate the applicability of our model on two tasks that have been widely studied - community detection and link prediction. We perform experiments on synthetic networks generated from ELSM and real world networks. Our model outperforms existing methods on standard benchmarks. While we report the results for both ELSM and iELSM in \S \ref{section:communitydetection} and \S \ref{section:linkprediction}, the inference procedure for ELSM has been described in supplementary material.

%========================================================================================================

\subsection{Synthetic Networks}
\label{section:generatingsyntheticnetworks}
We generated $10$ synthetic networks using Algorithm \ref{alg:generatingsyntheticdata}. For each network, we used $n=100$, $T=10$, $K=5$, $\mathbf{\pi}=[1/K, ..., 1/K]$, $\mathbf{m}=[0, 0]^\intercal$, $s=1.0$, $s_1=0.05$, $s_2=0.2$, $s_3=1.0$ and $s_4=0.5$ as input parameters. Community detection was performed on these networks using both normalized spectral clustering \cite{Luxburg:2007:ATutorialOnSpectralClustering} and our inference network as described in \S \ref{section:communitydetection}. The scores reported in Table \ref{table:clustering_results} for synthetic networks were obtained by averaging the scores across these $10$ networks. Visualization of some generated networks and the corresponding latent spaces has been provided in supplementary material. A video that shows the evolution of latent space over time has also been attached as part of supplementary material.

%========================================================================================================

\subsection{Real World Networks}
\label{section:realworldnetworks}
We evaluate our model on three real world dynamic networks apart from evaluating it on synthetic networks generated by ELSM. This section describes the datasets that we have used for our experiments:

\textbf{1) Enron email:}
The Enron dataset \cite{KlimtEtAl:2004:TheEnronCorpus} contains emails that were exchanged between 149 individuals over a period of three years. We use two different versions of this dataset - Enron-full and Enron-50. Enron-full has 12 snapshots, one for each month of the year 2002. In snapshot ${t}$, entry ${a_{ij}^{(t)}}$ counts the number of emails that were exchanged between users ${i}$ and ${j}$ during that month. We also consider unweighted Enron-full obtained by making all snapshots binary. Following \cite{FouldsEtAl:2011:ADynamicRelationalInfiniteFeatureModelForLongitudinalSocialNetworks}, Enron-50 considers only top 50 individuals with most number of emails across all snapshots. Each snapshot corresponds to a month (there are 37 snapshots). The adjacency matrix for each snapshot is symmetric and binary where ${a_{ij}^{(t)}} = 1$ if at least one email was exchanged between users ${i}$ and ${j}$ during that month and $0$ otherwise.

\textbf{2) NIPS co-authorship:}
There are 17 snapshots in this network, each corresponding to an year from 1987 to 2003. Entry ${a_{ij}^{(t)}}$ counts the number of NIPS conference publications in year ${t}$, that have individuals ${i}$ and ${j}$ as co-authors. We use a subset of this dataset - NIPS-110, that is created by making the snapshots binary and selecting top 110 authors based on the number of unique co-authors that they have had over these 17 years \cite{HeaukulaniEtAl:2013:DynamicProbabilisticModelsForLatentFeaturePropagationInSocialNetworks}.

\textbf{3) Infocom:} This dataset contains information about physical proximity between 78 individuals over a 93 hours long interval at Infocom 2006. Following \cite{KimEtAl:2013:NonparametricMultiGroupMembershipModelForDynamicNetworks}, we create snapshots corresponding to 1 hour long intervals. At time ${t}$, ${a_{ij}^{(t)}} = 1$ if both individuals ${i}$ and ${j}$ registered each others physical proximity during the ${t}^{th}$ hour. We remove those snapshots that have fewer than 72 non-zero entries which leaves us with 50 snapshots.

We next describe the two tasks that we have performed on various subsets of these datasets, namely - community detection and link prediction.

%========================================================================================================

\subsection{Community Detection}
\label{section:communitydetection}

\begin{table*}[t]
\caption{Community Detection Results: It can be seen that the detected communities are meaningful (as evident by modularity scores that are comparable to the ones obtained by the spectral clustering algorithm) while at the same time being \textit{smooth} (since the NMI scores are considerably higher) thereby validating our claim.}
\label{table:clustering_results}
\begin{center}
\begin{small}
\begin{sc}
\begin{tabular}{p{1.9cm}|ccc|ccc}
\toprule
\multirow{2}{*}{\textbf{Dataset}} & \multicolumn{3}{c|}{\textbf{Modularity}} & \multicolumn{3}{c}{\textbf{NMI}} \\
& Spectral & \textup{i}ELSM (Ours) & ELSM (Ours) & Spectral & \textup{i}ELSM (Ours) & ELSM (Ours) \\
\midrule
\textbf{Synthetic} & 0.479 & \textbf{0.489} & 0.488 & 0.769 & 0.851 & \textbf{0.864} \\
\textbf{Enron-full} \newline(Weighted)& 0.506 & \textbf{0.597} & 0.590 & 0.455 & 0.722 & \textbf{0.823} \\
\textbf{Enron-full} \newline(Binary)& 0.540 & \textbf{0.555} & 0.551 & 0.529 & 0.767 & \textbf{0.779} \\
\textbf{Enron-50}  & 0.396 & \textbf{0.419} & 0.414 & 0.560 & 0.819 & \textbf{0.838} \\
\textbf{NIPS-110}  & 0.497 & \textbf{0.601} & 0.595 & 0.249 & 0.804 & \textbf{0.863} \\
\textbf{Infocom}   & 0.283 & \textbf{0.288} & 0.270 & 0.443 & 0.643 & \textbf{0.662} \\
\bottomrule
\end{tabular}
\end{sc}
\end{small}
\end{center}
\end{table*}

The task is to assign each node to a community at each time step such that \textit{similar} nodes belong to the same community. To measure the quality of communities, at each time step, we use the well known modularity score. The modularity score lies in the range $[-1/2, 1)$, with values close to $1$ signifying \textit{good} communities (that have more edges within them as compared to the number of edges that will be obtained purely by chance).

We use the observed matrices $\{\mathbf{A}^{(1)}, \mathbf{A}^{(2)}, ..., \mathbf{A}^{(T)}\}$ to train the inference network for iELSM (\S \ref{section:networkarchitecture}) and ELSM (supplementary material). The latent embeddings obtained from the trained network are fed to the k-means clustering algorithm. At each time step, the optimal number of communities, $k^{(t)}$ is chosen from the range $[2, 10]$ by selecting the value of $k^{(t)}$ that maximizes the modularity score on the adjacency matrix that is induced by applying RBF kernel with variance parameter $1$ to the latent space embeddings. Note that we do not use the original adjacency matrix $\mathbf{A}^{(t)}$ while selecting the optimal number of communities.

We experiment with different variants of decoder in the inference network. For Enron-full, the edges have positive integer weights, we model this by imposing a Poisson distribution on $P({a_{ij}^{(t)}} | \mathbf{z}_{i}^{(t)}, \mathbf{z}_{j}^{(t)})$. Since iELSM (and ELSM) models the interaction between nodes as a function of distance between their latent embeddings, we predict the mean of Poisson distribution for position $({i}, {j})$ at time step ${t}$ as:
\begin{equation}
	\label{eq:poissonmeanprediction}
	\rho_{ij}^{(t)} = \exp (-w_\rho^2 ||\mathbf{z}_{i}^{(t)} - \mathbf{z}_{j}^{(t)}||^2 + b_\rho)
\end{equation} 
The parameters $w_\rho$ and $b_\rho$ are shared across all time steps and positions. These parameters are learned using standard backpropagation ((\ref{eq:poissonmeanprediction}) represents a single layer, single node neural network with $\exp(.)$ as the activation function). One can also employ a similar technique for learning parameters ${s_2}$ (\S \ref{section:generatingobservedatfromlatentzt}) and ${s_4}$ (\S \ref{section:evolvingthelatentspace}) from data by using a single layer, single node neural network to learn optimal scaling of distances in (\ref{eq:observedgivenlatent}) and (\ref{eq:latentupdatemean}).

The key claim that we wish to demonstrate here is that the latent embeddings learned by our model are not only good for finding communities at individual time steps, but also lead to a more plausible, gradually changing community structure over time. To do so, we compute the NMI score between community assignments at successive time steps. The NMI score lies in $[0, 1]$ with values close to $1$ signifying gradual change in community structure in our setup. 

We compare our scores against the scores obtained by independently applying normalized spectral clustering \cite{Luxburg:2007:ATutorialOnSpectralClustering} at each snapshot. The number of communities, $k^{(t)}$, is chosen by using the same method that was used for our model on the node embeddings produced by spectral clustering. 

Table \ref{table:clustering_results} summarizes the average modularity and NMI scores for various datasets. Although  $k^{(t)}$ is chosen based on the graph induced by latent embeddings, the scores reported in Table \ref{table:clustering_results} correspond to the observed graph for the chosen number of communities. From Table \ref{table:clustering_results}, it is clear that our model achieves modularity scores that are at par (or better) with respect to spectral clustering while exhibiting a significantly higher NMI score, thus validating our claim.

%========================================================================================================

\subsection{Link Prediction}
\label{section:linkprediction}

\begin{table}[t]
\caption{Link Prediction Results: Our method outperforms other approaches on both metrics. We were unable to obtain an implementation for DMMG and hence the performance numbers of DMMG on Enron-50 are missing.}
\label{table:linkpredictionresults}
\begin{center}
\begin{small}
\begin{sc}
\begin{tabular}{p{1.1cm}C{0.7cm}C{0.7cm}C{0.7cm}C{0.7cm}C{0.7cm}C{0.7cm}}
\toprule
\multirow{2}{*}{} & \multicolumn{2}{c}{\textbf{Enron-50}} & \multicolumn{2}{c}{\textbf{Infocom}} & \multicolumn{2}{c}{\textbf{NIPS-110}} \\
& AUC & F1  & AUC & F1 & AUC & F1 \\
\midrule
BAS  & 0.874 & 0.585 & 0.698 & 0.317 & 0.703 & 0.161 \\
LFRM & 0.777 & 0.312 & 0.640 & 0.248 & 0.398 & 0.011 \\
DRIFT& 0.910 & 0.578 & 0.782 & 0.381 & 0.672 & 0.084 \\
DMMG & - & - & 0.804 & 0.392 & 0.732 & 0.196 \\
\textup{i}ELSM (Ours) & 0.913 & 0.600 & 0.868 & \textbf{0.489} & \textbf{0.754} & 0.248 \\
ELSM (Ours) & \textbf{0.911} & \textbf{0.596} & \textbf{0.871} & \textbf{0.489} & 0.742 & \textbf{0.251} \\
\bottomrule
\end{tabular}
\end{sc}
\end{small}
\end{center}
\end{table}

Good performance on community detection task testifies that the latent embeddings being learned by the network have a nice structure. But how does one ensure that the modeled network dynamics are faithful to real world data? Is our model overfitting to the training data to get good embeddings? These questions get answered, if the observed data can be extrapolated to predict an unseen network snapshot using our model. This task is known as link prediction.

Formally, given network snapshots up to time ${t}$, $\{\mathbf{A}^{(1)}, \mathbf{A}^{(2)}, ..., \mathbf{A}^{(t)}\}$, we want to predict the next network snapshot $\mathbf{A}^{(t+1)}$. Note that this involves predicting the appearance of new links as well as removal of existing links. We use binary networks in this experiment to compare against other approaches. We use the well known AUC score and F1 score for the purpose of  comparison. Values close to $1$ indicate good performance for both the scores.

To predict $\mathbf{A}^{(t+1)}$, we train the inference network for iELSM and ELSM on $\{\mathbf{A}^{(1)}, \mathbf{A}^{(2)}, ..., \mathbf{A}^{(t)}\}$. We update the predicted latent embeddings $\mathbf{Z}^{(t)}$ using (\ref{eq:latentupdatemean}) and (\ref{eq:updatedembedding}) to get  $\mathbf{Z}^{(t + 1)}$. Then, $P({a_{ij}}^{(t+1)} = 1 | \mathbf{Z}^{(t + 1)})$ is computed using (\ref{eq:observedgivenlatent}) (or the decoder, if a decoder network has been used). Note that for ELSM, while updating the latent embeddings we set $h_i^{(t+1)} = 0$ for all nodes.

We compare our performance against a simple baseline (BAS) in which the probability of an edge is directly proportional to the number of times it has been observed in the past \cite{FouldsEtAl:2011:ADynamicRelationalInfiniteFeatureModelForLongitudinalSocialNetworks}. We also compare against existing approaches - LFRM \cite{MillerEtAl:2009:NonparametricLatentFeatureModelsForLinkPrediction} (using only the last snapshot), DRIFT \cite{FouldsEtAl:2011:ADynamicRelationalInfiniteFeatureModelForLongitudinalSocialNetworks} and DMMG \cite{KimEtAl:2013:NonparametricMultiGroupMembershipModelForDynamicNetworks} (Table \ref{table:linkpredictionresults}). Maximum F1 score over all thresholds is selected at each snapshot as it was done in \cite{KimEtAl:2013:NonparametricMultiGroupMembershipModelForDynamicNetworks}. The scores reported here have been obtained by averaging the snapshot wise scores. It can be seen that our method outperforms other methods on both metrics. Visualization of latent embeddings and predicted output matrices for Enron-full can be found in supplementary material.

%%%%%%%%%%%%%%%%%%%%%%%%%%%%%%%%%%%%%%%%%%%%%%%%%%%%%%%%%%%%%%%%%%%%%%%%%%%%%%%%%%%%%%%%%%%%%%%%%%%%%%%%%%%%%%%%%%%%%%%%%%%%%%

\section{Conclusion and Future Work}
\label{section:conclusionandfuturework}
In this paper, we proposed ELSM, a generative model for dynamically evolving networks. We also proposed a neural network architecture that performs approximate inference in a simplified version of our model, iELSM (inference for ELSM is in supplementary material) and highlighted the flexibility of this approach. Our model is capable of: (i) Generating synthetic dynamic networks with gradually evolving communities and (ii) Learning meaningful latent embeddings of nodes in a dynamic network. We demonstrated the quality of learned latent embeddings on downstream tasks like community detection and link prediction in dynamic networks.

In this paper we focused on undirected, positive influence based, gradually changing, assortative networks with a fixed number of nodes. Though, these properties are exhibited by a large number of real world networks, however there are other important classes of networks that do not follow these properties. For example, one can also extend this idea to directed networks. One can also consider the case where the number of nodes is allowed to change over time. Considering networks that are not necessarily assortative (like hierarchical networks) also poses interesting questions.

%%%%%%%%%%%%%%%%%%%%%%%%%%%%%%%%%%%%%%%%%%%%%%%%%%%%%%%%%%%%%%%%%%%%%%%%%%%%%%%%%%%%%%%%%%%%%%%%%%%%%%%%%%%%%%%%%%%%%%%%%%%%%%

\bibliography{Bibliography-File}
\bibliographystyle{aaai}

\newpage
\appendix
\section{Inference in Evolving Latent Space Model}
\label{appendix:inferenceinfullELSM}

In this section we describe the inference in ELSM. The objective is to infer a distribution over all the other variables in the model given the observations $\{\mathbf{A^{(1)}}, .., \mathbf{A^{(T)}}\}$. The expression for joint log probability of all variables can be computed from the graphical model presented in Figure \ref{fig:graphicalmodel} in a similar manner as it was done in (\ref{eq:sELSMjointloglikelihood}):
\begin{equation}
    \label{eq:ELSMjointlikelihood}
    \begin{split}
    \log P(\mathbf{A}^{(1)}, .., \mathbf{A}^{(T)}, \mathbf{Z}^{(1)}, ..., \mathbf{Z}^{(T)}, {\mathbf{h}^{(2)}}, ..., {\mathbf{h}^{(T)}}, \bm{\alpha}^{(2)}, \\..., \bm{\alpha}^{(T)}, \mathbf{c}, \bm{\mu}_1, ..., \bm{\mu}_K) = 
    \sum_{i=1}^{n} \log P({c_i}) + 
    \sum_{j=1}^{K} \log P(\bm{\mu}_j) \\ +
    \sum_{t=2}^{T} \log P(\bm{\alpha}^{(t)}) + 
    \sum_{t=1}^{T} \sum_{i > j} \log P({a_{ij}^{(t)}} | \mathbf{z}_i^{(t)}, \mathbf{z}_j^{(t)}) + \\
    \sum_{i=1}^{n} \log P(\mathbf{z}_{i}^{(1)} | {c_i}, \bm{\mu}_{c_i}) +
    \sum_{t=2}^{{T}} \sum_{i=1}^{n} \log P({h_{i}^{(t)}} | \mathbf{z}_i^{(t-1)}, \bm{\alpha}^{(t)}) \\+
    \sum_{t=2}^{{T}} \sum_{i=1}^{n} \log P(\mathbf{z}_{i}^{(t)} | \mathbf{Z}^{(t-1)}, \mathbf{A}^{(t-1)}, {h_i^{(t)}}, \bm{\alpha}^{(t)})
    \end{split}
\end{equation}
As in (\ref{eq:meanfield}), we define a distribution $Q(., \theta)$ that belongs to the mean-field family. As before, $\theta$ represents parameters of the underlying neural network that parameterizes the variational distribution $Q(., \theta)$. The distribution $Q(., \theta)$ can be factored as:
\begin{equation}
	\label{eq:ELSMmeanfield}
	\begin{split}
	Q(\mathbf{Z}^{(1)}, ..., \mathbf{Z}^{(T)}, {\mathbf{h}^{(2)}}, ..., {\mathbf{h}^{(T)}}, \bm{\alpha}^{(2)}, ..., \bm{\alpha}^{(T)}, \mathbf{c}, \bm{\mu}_1, \\..., \bm{\mu}_K) = 
	\Big(\prod_{i=1}^{n} q^{c}_i({c_i})\Big)
	\Big(\prod_{j=1}^{K} q^{\mu}_j(\bm{\mu}_j)\Big)
	\Big(\prod_{t=2}^{T} q^{\alpha}_t(\bm{\alpha}^{(t)})\Big) \\
	\Big(\prod_{t=2}^{T} \prod_{i=1}^{n} q^{h}_{ti}({h_i^{(t)}})\Big)
	\Big(\prod_{t=1}^{T} \prod_{i=1}^{n} q^{z}_{ti}(\mathbf{z}_i^{(t)})\Big)	
	\end{split}
\end{equation}
Here, the dependence of individual factors on $\theta$ is implicit and has been avoided for the sake of clarity. We model each of these factors as follows:
\begin{equation}
	\label{eq:c_factor}
	q_i^c \sim Multinomial(\mathbf{\hat{c}}_i; \theta)
\end{equation}
\begin{equation}
	\label{eq:mu_factor}
	q_j^\mu \sim \mathcal{N}(\mathbf{m}_{\mu_j}, \mathbf{s}_{\mu_j}^2 \mathbf{I}; \theta)
\end{equation}
\begin{equation}
	\label{eq:alpha_factor}
	q_t^\alpha \sim \mathcal{N}(\mathbf{m}^{(t)}_{\alpha}, (\mathbf{s}^{(t)}_{\alpha})^2 \mathbf{I}; \theta)
\end{equation}
\begin{equation}
	\label{eq:h_factor}
	q_{ti}^h \sim Bernoulli(\hat{h}^{(t)}_{i}; \theta)
\end{equation}
\begin{equation}
	\label{eq:z_factor}
	q_{ti}^z \sim \mathcal{N}(\bm{\nu}_i^{(t)}, (\bm{\sigma}_i^{(t)})^2\mathbf{I}; \theta)
\end{equation}

The variational parameters $\mathbf{\hat{c}}_i$, $\mathbf{m}_{\mu_j}$, $\mathbf{s}_{\mu_j}^2$, $\mathbf{m}^{(t)}_{\alpha}$, $(\mathbf{s}^{(t)}_{\alpha})^2$, $\hat{h}^{(t)}_{i}$, $\bm{\nu}_i^{(t)}$ and $(\bm{\sigma}_i^{(t)})^2$ are predicted by the inference network which has been shown in Figure \ref{fig:ELSMnetworkmodel}. The objective of this network is to maximize $\mathrm{ELBO}$ for ELSM. $\mathrm{ELBO}$ can be computed by plugging in (\ref{eq:ELSMjointlikelihood}) and (\ref{eq:ELSMmeanfield}) into the expression for $\mathrm{ELBO}$ in the same way as it was done in (\ref{eq:elbo}).

We compute the expectations over ${h_i^{(t)}}$ and ${c_i}$, ${i} = 1, 2, ..., {n}$, ${t} = 1, 2, .., {T}$ analytically for the joint log likelihood term in $\mathrm{ELBO}$. The entropy terms can also be evaluated analytically using standard formulas for Gaussian, Bernoulli and Multinomial distributions. For all other terms, we approximate the expectation via Monte-Carlo method using only one sample.

%=======================================================================================================

\subsection{Network Architecture}
\label{appendix:ELSMnetworkarchitecture}
A bidirectional LSTM is at the core of our inference network for ELSM. At each timestep, it takes the observed adjacency matrix $\mathbf{A}^{(t)}$ as input. As in case of iELSM, each row of $\mathbf{A}^{(t)}$ is treated as a feature vector for the corresponding node. For each node, the hidden state of LSTM is updated via the normal LSTM update rules to generate $\mathbf{g}_i^{(t)}$, $i = 1, 2, ..., {n}$, $t = 1, 2, ..., {T}$. All the variational parameters are calculated based on the hidden state.

\begin{figure}
%\vskip 0.2in
\begin{center}
\centering
\includegraphics[width=\columnwidth]{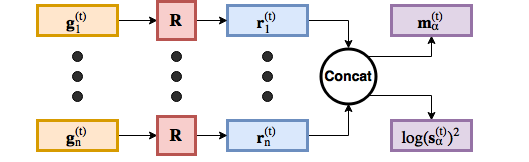}
\caption{$\bm{\alpha}$ Network - $\mathbf{P}$ is a sub-network that reduces the dimension of input before concatenation.}
\label{fig:ELSMalphanetwork}
\end{center}
%\vskip -0.2in
\end{figure}

Since multiple nodes come together to form a new community, calculation of $\mathbf{m}^{(t)}_{\alpha}$ and $(\mathbf{s}^{(t)}_{\alpha})^2$ requires information about all nodes in the graph. At timestep $t$, the hidden states for all nodes $\mathbf{g}_i^{(t)}$, $i = 1, 2, ..., {n}$, are passed to network $\bm{\alpha}$ which is shown in Figure \ref{fig:ELSMalphanetwork} to compute $\mathbf{m}^{(t+1)}_{\alpha}$ and $(\mathbf{s}^{(t+1)}_{\alpha})^2$. The $\bm{\alpha}$ network performs two operations: (i) it reduces the dimension of $\mathbf{g}_i^{(t)}$ to obtain $\mathbf{r}_i^{(t)}$ via a non-linear layer $\mathbf{R}$ that is shared across all nodes and timesteps and (ii) it concatenates $\mathbf{r}_i^{(t)}$, $i = 1, 2, ..., {n}$ and obtains $\mathbf{m}^{(t+1)}_{\alpha}$ and $(\mathbf{s}^{(t+1)}_{\alpha})^2$ by transforming the concatenated vector. 

\begin{figure*}
%\vskip 0.2in
\begin{center}
\centering
\includegraphics[width=\textwidth]{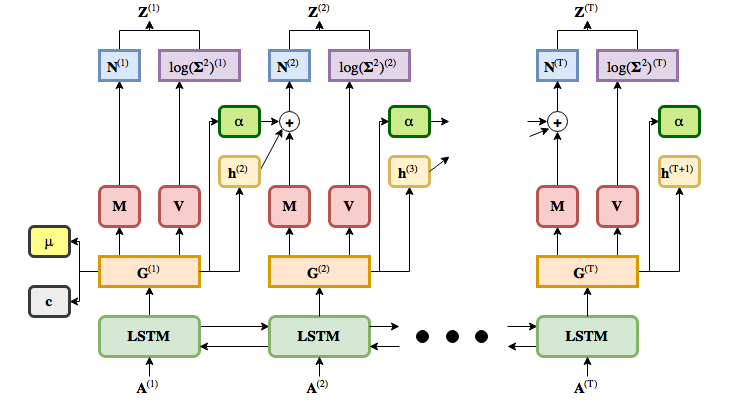}
\caption{Time Unrolled Inference Network for ELSM. $\mathbf{M}$ and $\mathbf{V}$ are neural networks for predicting mean and $\log$-variance parameters as in the case of iELSM. Additionally, we have four extra modules. The network $\bm{\alpha}$ predicts $\bm{\alpha}^{(t + 1)}$ at timestep $t$. In the first timestep, the network $\bm{\mu}$ predicts the initial community centers and $\mathbf{c}$ predicts the initial membership to communities. At timestep $t$, another module (not explicitely shown in the figure) predicts $h^{(t + 1)}_i$ for all nodes $i$. All these additional network modules take the LSTM hidden state as input.}
\label{fig:ELSMnetworkmodel}
\end{center}
%\vskip -0.2in
\end{figure*}

\begin{figure}
%\vskip 0.2in
\begin{center}
\centering
\includegraphics[width=\columnwidth]{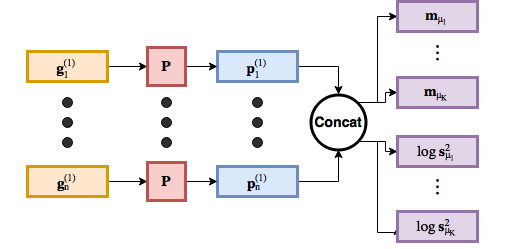}
\caption{$\bm{\mu}$ Network - $\mathbf{R}$ is a sub-network that reduces the dimension of input before concatenation.}
\label{fig:ELSMmunetwork}
\end{center}
%\vskip -0.2in
\end{figure}

Similar to the case of $\bm{\alpha}^{(t)}$, calculation of $\mathbf{m}_{\mu_j}$ and $\mathbf{s}_{\mu_j}^2$ for the initial community centers $\bm{\mu}_j$, $j = 1, 2, ..., {K}$, requires information about all nodes. The network $\bm{\mu}$ in Figure \ref{fig:ELSMnetworkmodel} (explained in Figure \ref{fig:ELSMmunetwork}) performs the task of computing  $\mathbf{m}_{\mu_j}$ and $\mathbf{s}_{\mu_j}^2$. The architecture of this network is similar to the architecture of $\bm{\alpha}$ network that has been described above.

In the first layer, the latent space comes from a Gaussian mixture model with means given by $\bm{\mu}_j$, $j = 1, 2, ..., {K}$. Once the latent embedding of each node and the means $\bm{\mu}_j$ have been predicted by the network, $\mathbf{\hat{c}}_i$ can be derived in the same way as it is done in a Gaussian mixture model, i.e. for each $j = 1, 2, ..., {K}$,
\begin{equation}
	\label{eq:ciELSM}
	(\hat{c}_i)_j = \frac{\mathcal{N}(\mathbf{z}_i^{(1)}\, |\, \bm{\mu}_j,\, {s_4^2}\mathbf{I}) P({c_i} = j)}{\sum_{k=1}^{{K}} \mathcal{N}(\mathbf{z}_i^{(1)}\, |\, \bm{\mu}_k,\, {s_4^2}\mathbf{I}) P({c_k} = j)}  
\end{equation} 

The parameter $\hat{h}^{(t+1)}_{i}$ in (\ref{eq:h_factor}) is predicted by passing $\mathbf{g}_i^{(t)}$ through a linear layer and applying the sigmoid activation to the output. Since we have a bidirectional LSTM, the  network can encode information about switching of a node to a new community at time $t + 1$ in $\mathbf{g}_i^{(t)}$. This allows us to predict $\hat{h}^{(t+1)}_{i}$ from $\mathbf{g}_i^{(t)}$.

Finally, after predicting $\hat{h}_{ti}$ and $\bm{\alpha}^{(t)}$, we predict the mean $\mathbf{m}_i^{(t)}$ for each node using the network $\mathbf{M}$ in Figure \ref{fig:ELSMnetworkmodel} that takes $\mathbf{g}_i^{(t)}$ as input to produce $\mathbf{m}_i^{(t)}$. The predicted $\mathbf{m}_i^{(t)}$ can be thought of as the latent embedding of node ${i}$ at timestep ${t}$ if the splitting procedure were absent. The parameter $\bm{\nu}_i^{(t)}$ can now be modeled as a convex combination of $\mathbf{m}_i^{(t)}$ and $\bm{\alpha}^{(t)}$:
\begin{equation}
	\label{eq:nuitELSM}
	\bm{\nu}_i^{(t)} = (1-\hat{h}^{(t)}_{i}) \mathbf{m}_i^{(t)} + \hat{h}^{(t)}_{i}\bm{\alpha}^{(t)}, \,\,\,\, t = 2, 3, ..., {T}
\end{equation}
Note that (\ref{eq:nuitELSM}) is valid only for $t = 2, 3, ..., {T}$. For $t=1$, splitting does not happen and hence $\bm{\nu}_i^{(1)} = \mathbf{m}_i^{(1)}$. The network $\mathbf{V}$ in Figure \ref{fig:ELSMnetworkmodel} predicts $\log (\bm{\sigma}_i^2)^{(t)}$ in the same was as it was done in iELSM.  Based on $\bm{\nu}_i^{(t)}$ and $\log (\bm{\sigma}_i^2)^{(t)}$, the latent embeddings can be sampled.

\section{Generated Synthetic Networks}
\label{appendix:generatedsyntheticnetworks}

In this section, we show an example network that was generated using ELSM along with the corresponding latent embeddings of nodes at each time step. Synthetic dynamic networks with desired properties can be generated using ELSM by setting various input attributes in Algorithm \ref{alg:generatingsyntheticdata} appropriately. As an example of varying the input parameters, we explore the effect of changing parameter $s_2$ on sparsity of the generated network.

We generated a synthetic network with $n=100$ nodes and $T=5$ snapshots using Algorithm \ref{alg:generatingsyntheticdata}. Other input parameters were set as follows: $K=5$, $\mathbf{\pi}=[1/K, ..., 1/K]$, $\mathbf{m}=zeros(2)$, $s=1.0$, $s_1=0.05$, $s_2=0.2$, $s_3=1.0$ and $s_4=1.5$. Note that we used a two dimensional latent space so that the latent embeddings can be plotted easily.

\begin{figure*}
\vskip 0.2in
\begin{center}
\centering
\includegraphics[width=0.8\textwidth]{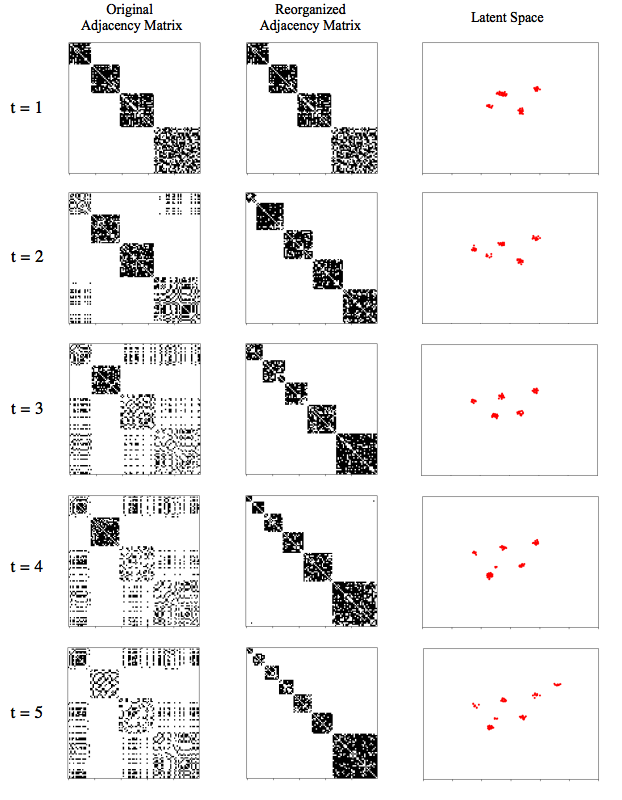}
\caption{Generated Synthetic Network - It can be seen that community structure is maintained over time. Moreover, new communities take birth and older ones die in the process. See the accompanying video for another example.}
\label{fig:generatednetwork}
\end{center}
\vskip -0.2in
\end{figure*}

Each row in Figure \ref{fig:generatednetwork} corresponds to one time step in the network. The center subplot in each row shows the adjacency matrix which has been reorganized to make the community structure in the network visible. The left subplot freezes the position of nodes in adjacency matrix to show how nodes interact with other nodes over time. Note that in the first row, the left and center subplots are same and thereafter the same ordering of nodes in left subplot is used for all subsequent time steps. The subplot on right shows the latent embeddings of all the nodes.

\begin{figure*}
%\vskip 0.2in
\begin{center}
\centering\includegraphics[width=0.6\textwidth]{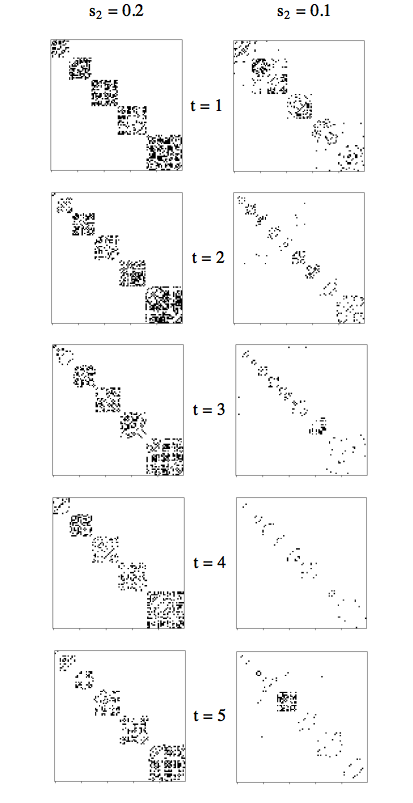}
\caption{Effect of $s_2$ on Sparsity of Generated Network}
\label{fig:generatednetworksparse}
\end{center}
%\vskip -0.2in
\end{figure*}

In Figure \ref{fig:generatednetworksparse}, we show the observed adjacency matrices of two dynamic networks. Both the networks were generated using ELSM. The first network used the following set of input parameters for Algorithm \ref{alg:generatingsyntheticdata}: $n=100$, $T=5$, $K=5$, $\mathbf{\pi}=[1/K, ..., 1/K]$, $\mathbf{m}=zeros(2)$, $s=1.0$, $s_1=0.1$, $s_2=0.2$, $s_3=1.0$ and $s_4=1.5$. The second network was also generated using the same set input parameters except for $s_2$ that was set to $0.1$, thus making it more sparse.

Each row in Figure \ref{fig:generatednetworksparse} corresponds to one time step. In each row, the subplot on left shows adjacency matrix of the first network and the subplot on right shows adjacency matrix of the second network. One can clearly see that the second network is more sparse as compared to the first network. Other input parameters can be similarly modified to get networks with different properties.

We created a video that shows the evolution of latent embeddings in a synthetically generated dynamic network over time. It has been attached as part of supplementary material. This video was generated by taking $n=200$, $T=200$, $K=5$, $\mathbf{\pi}=[1/K, ..., 1/K]$, $\mathbf{m}=zeros(2)$, $s=1.0$, $s_1=0.05$, $s_2=0.5$, $s_3=1.0$ and $s_4=1.5$.

\section{Other Visualizations}
\label{appendix:othervisualizations}
To demonstrate the working of our model on a real world network, we will visualize the Enron-full dataset described in Section \ref{section:realworldnetworks} along with the learned latent embeddings. 

\begin{figure*}
\vskip 0.2in
\begin{center}
\centering
\includegraphics[scale=0.50]{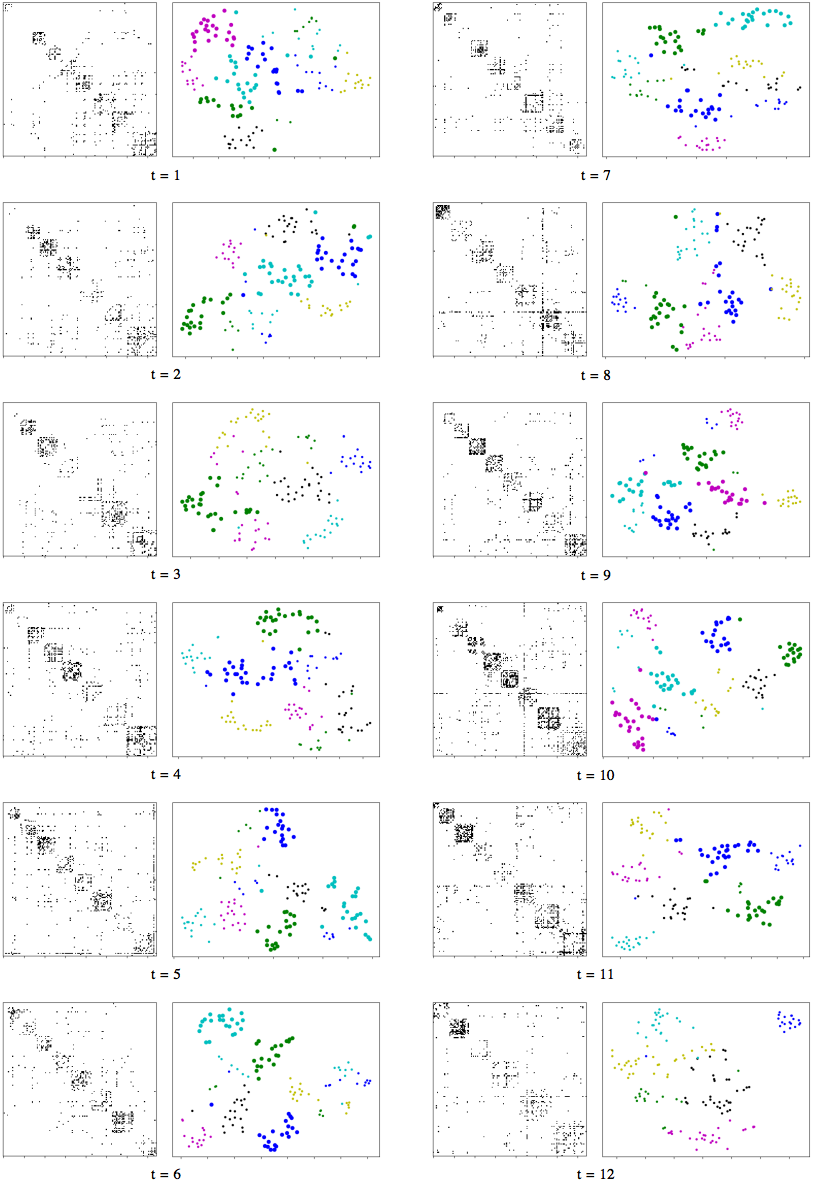}
\caption{Visualizing Latent Embeddings for Enron-full. It can be seen that the latent space possesses a community structure. This is because all operations in our model are dependent on the euclidean distance between latent embeddings.}
\label{fig:enron_clustering}
\end{center}
\vskip -0.2in
\end{figure*}

We trained an iELSM inference network on this dataset after making all snapshots binary. The embeddings learned by this network (we used $d=8$) have been plotted by using t-SNE \cite{Maaten:2008:VisualizingDataUsingTSNE} in Figure \ref{fig:enron_clustering}. Each row corresponds to two time steps. The first two columns show $\mathbf{A}^{(t)}$ and the t-SNE plot corresponding to $\mathbf{Z}^{(t)}$ while the next two columns show $\mathbf{A}^{(t+6)}$ and the t-SNE plot corresponding to $\mathbf{Z}^{(t+6)}$ for $t = 1, 2, ..., 6$.

\begin{figure*}
\vskip 0.2in
\begin{center}
\centering
\includegraphics[scale=0.60]{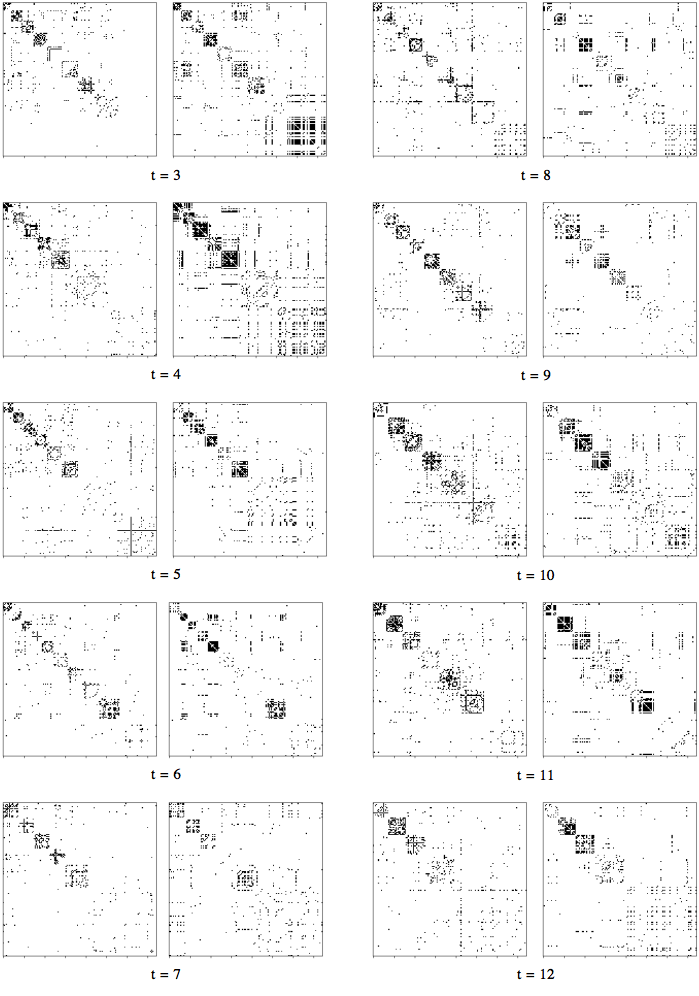}
\caption{Visualizing Original and Predicted Matrices for Enron-full. For each timestep, the figure on the left corresponds to the original adjacency matrix and the figure on the right corresponds to the predicted adjacency matrix.}
\label{fig:enron_link_pred}
\end{center}
\vskip -0.2in
\end{figure*}

We trained another iELSM inference network for link prediction on Enron-full to get the edge prediction probabilities. At each time step, we selected a threshold that maximized the F1 score for that snapshot. Based on the selected threshold, we obtained a binary prediction matrix $\hat{\mathbf{A}}^{(t)}$ at each time step. Figure \ref{fig:enron_link_pred} shows observed $\mathbf{A}^{(t)}$ alongside the predicted $\mathbf{\hat{A}}^{(t)}$ for $t = 3, 4, ..., T$. In Figure \ref{fig:enron_link_pred}, for each time step $t$, we have rearranged the elements of observed $\mathbf{A}^{(t)}$ to make the community structure clear. We use the same arrangement of nodes in $\mathbf{\hat{A}}^{(t)}$ that was used for the corresponding $\mathbf{A}^{(t)}$.

\end{document}